\documentclass{PoS}

\usepackage{xspace}

\def\xmax{\ensuremath{X_{\rm max}}\xspace}

\def\nmu{\ensuremath{N_\mu}\xspace}

\usepackage{units}
\usepackage{paralist}
\usepackage{cite}

\title{The hadronic interaction model Sibyll 2.3c and Feynman scaling}

\ShortTitle{The hadronic interaction model Sibyll}

\author{Felix Riehn\\
       LIP Lisbon, Av. Prof. Gama Pinto 2, 1649-003 Lisbon, Portugal\\
       E-mail: \email{friehn@lip.pt}}

\author{Hans P. Dembinski\\ Max Planck Institute for Nuclear Physics,
  Postfach 103980, 69029 Heidelberg, Germany }

\author{\speaker{Ralph Engel}\\
  Karlsruher Institut f\"ur Technologie, Institut f\"ur Kernphysik, Postfach 3640, 76021 Karlsruhe, Germany}

\author{Anatoli Fedynitch\\
  DESY, Platanenallee 6, 15738 Zeuthen, Germany
}

\author{Thomas K. Gaisser\\
  Bartol Research Institute, Department of Physics and Astronomy, University of Delaware, Newark, DE 19716, USA}

\author{Todor Stanev\\
  Bartol Research Institute, Department of Physics and Astronomy, University of Delaware, Newark, DE 19716, USA}

\abstract{The Monte Carlo model Sibyll has been designed for efficient 
  simulation of hadronic multiparticle production up to the highest
  energies as needed for interpreting cosmic ray measurements.
  For more than 15 years, version 2.1 of Sibyll has been one of the standard models for air shower simulation.
  Motivated by 
  data of LHC and fixed-target experiments and a better understanding
  of the phenomenology of hadronic interactions, we have developed an improved
  version of this model, version 2.3, which has been released in 2016. In this contribution 
  we present a revised version of this model, called Sibyll 2.3c, that is
  further improved by adjusting particle production spectra to match
  the expectation of Feynman scaling in the fragmentation region. After a brief introduction to
  the changes implemented in Sibyll 2.3 and 2.3c with respect to Sibyll 2.1, the current
  predictions of the model for the depth of shower maximum, the number of
  muons at ground, and the energy spectrum of muons in extensive air showers are
  presented.
}

\FullConference{35th International Cosmic Ray Conference\\
         10-20 July, 2017\\
         Bexco, Busan, Korea}

\begin{document}

\section{Introduction}

\noindent
Sibyll is one of the standard hadronic event generators used in the simulation
of extensive air showers initiated by high energy cosmic rays. It is
designed to describe the general features of hadronic multiparticle
production, like the leading particle effect, the 
formation of high-$p_{\rm T}$ jets predicted in
QCD, the production of diffractively excited states of the projectile and target,
and approximate scaling of leading particle distributions with interaction
energy. In Sibyll, focus is put on those physics aspects that are most relevant for the
development of extensive air showers, like energy flow and particle production in 
the forward phase space region. While the model is kept as
simple as possible, the important microscopic physics concepts
and the general principles of scattering theory and unitarity are implemented to allow
extrapolation to energies and phase space regions beyond the reach of
colliders~\cite{Engel:1992vf,Fletcher:1994bd,Ahn:2009wx}.

In the recently released Sibyll~2.3~\cite{Engel:2015dxa} the previous
model version~\cite{Ahn:2009wx} was
extended and updated while retaining the overall underlying picture of hadronic
interactions implemented already in Sibyll 2.1. The most notable changes of Sibyll~2.3 wrt.{} version 2.1 are
\noindent
\begin{compactitem}
\item
  New fits to total and elastic cross sections for $p$-$p$, $\pi$-$p$
  and $K$-$p$ interactions to match new LHC and fixed-target data.
\item
  Explicit treatment of remnant excitations for better description of leading particle production.
\item
  Enhanced production of vector resonances in the fragmentation region
  of mesons ($\rho^0$).
\item
  Implementation of diffraction dissociation in interactions of hadrons with nuclei based on 
  a two-component model (ground state and excited state of projectile and target hadrons),
  similar to the Good-Walker model of diffraction~\cite{Good:1960ba}.
\item
  Increase of the rate of baryon-antibaryon pair production in string fragmentation,
  including a higher production rate in minijet fragmentation than in soft processes.
\item
  Implementation of a phenomenological model for describing the production of charm particles.
\end{compactitem}
A summary of the physics ideas and principles on which Sibyll is based can be found in Ref.~\cite{Engel:2017wpa}
and a first description of Sibyll 2.3 is given in Ref.~\cite{Engel:2015dxa}.
Here we will discuss the update of Sibyll~2.3 to 2.3c.

\section{Scaling behavior of leading particle distributions and extrapolation to high energy}

\noindent
One of the important tasks of interaction models used in astroparticle
physics is the extrapolation from the energy range and phase space
covered by laboratory measurements to energies and phase space regions that
play a role in air shower development~\cite{Engel:2011zzb}
or in the production of atmospheric leptons~\cite{Gaisser:2002jj}.
While measurements at LHC, for example,
extend to $\sim 10\,$TeV in cm.\ energy, the interaction energies of cosmic rays, as
they are studied by the Pierre Auger Observatory, extend up to
cm.\ energies of $400\,$TeV. In terms of phase space the situation is
worse still. Interactions
with small momentum transfers (low-$Q^2$, 'soft' scattering) are most
common. These interactions cannot be calculated within perturbative QCD,
the currently best approach for quantitatively predicting hadronic
particle production. Moreover, in such interactions, particles are emitted at very small
angles, which  means they escape
detection through the beam pipe in particle collider experiments.


\subsection{Feynman scaling and leading particles}

\noindent
To allow for meaningful predictions, additional principles have to be
invoked. One such principle is that of Feynman scaling~\cite{Feynman:1969ej}. It
states that the production cross section of particle $i$ at high energy is
independent of the interaction energy and only depends on transverse
momentum and the longitudinal momentum fraction $x_{\rm
  F}=p_{z}/p^{\rm max}_{z}$
\begin{equation}
  E \frac{\mathrm{d}^3\sigma_i}{\mathrm{d}^3\vec{p}} \to f_i(p_{\rm T},x_{\rm F}).
\end{equation}
While Feynman scaling was quickly found to be violated by jet
formation in hard scattering processes~\cite{Alner:1986xu}, it is
still believed to hold approximately in the so-called fragmentation region where
soft interactions dominate. Translated into longitudinal phase space,
one expects a universal shape for the production cross section of a given particle 
for $x_{\rm F}\gtrsim 0.1$ at high energy.  
Assuming Feynman scaling, the production spectra in
an interaction model can be tuned to measurements at low energy where
full phase space coverage is still feasible and then reliably
extrapolated to high energies.

In parton-based interaction models such as Sibyll, Feynman scaling
is not imposed on the hadron level. Instead, the momentum distributions of the partons
at the string ends and, correspondingly, of the hadrons produced in string fragmentation
satisfy Feynman scaling. Due to overall energy-momentum conservation
and the rapid increase of the secondary particle multiplicity,
violation of Feynman scaling of leading particles is ultimately expected at very high energy.
It is an open question to what extent Feynman scaling might be violated
for leading particles at intermediate energies. In the limit of black disc scattering even a total
absence of a leading particle effect, and hence maximum violation of Feynman
scaling of the leading particles, can be expected~\cite{Drescher:2004sd}.

Sibyll~2.3c has been tuned to represent the most conservative assumption,
namely minimal violation of Feynman scaling of leading particle distributions.
Other models (EPOS~\cite{Werner:2005jf}, QGSjet~\cite{Ostapchenko:2010vb,Ostapchenko:2016ytp})
show a stronger violation of Feynman scaling.

\subsection{Sibyll predictions for leading particle distributions}

\begin{figure*}[tb!]
  \includegraphics[width=0.5\columnwidth]{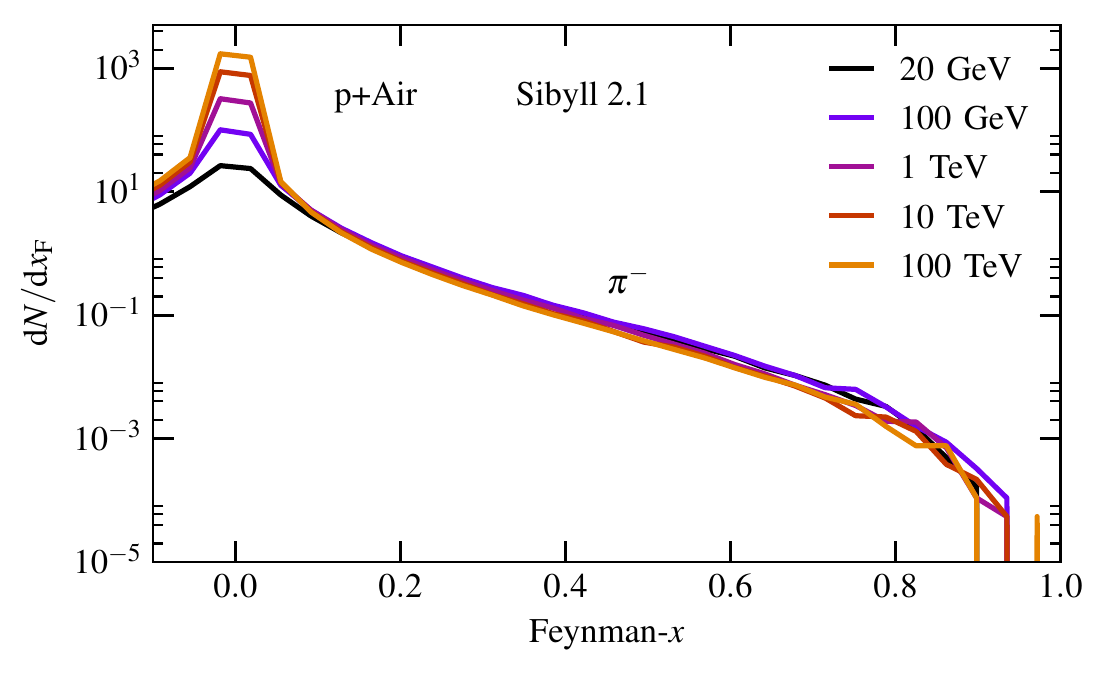}
  \hfill
  \includegraphics[width=0.5\columnwidth]{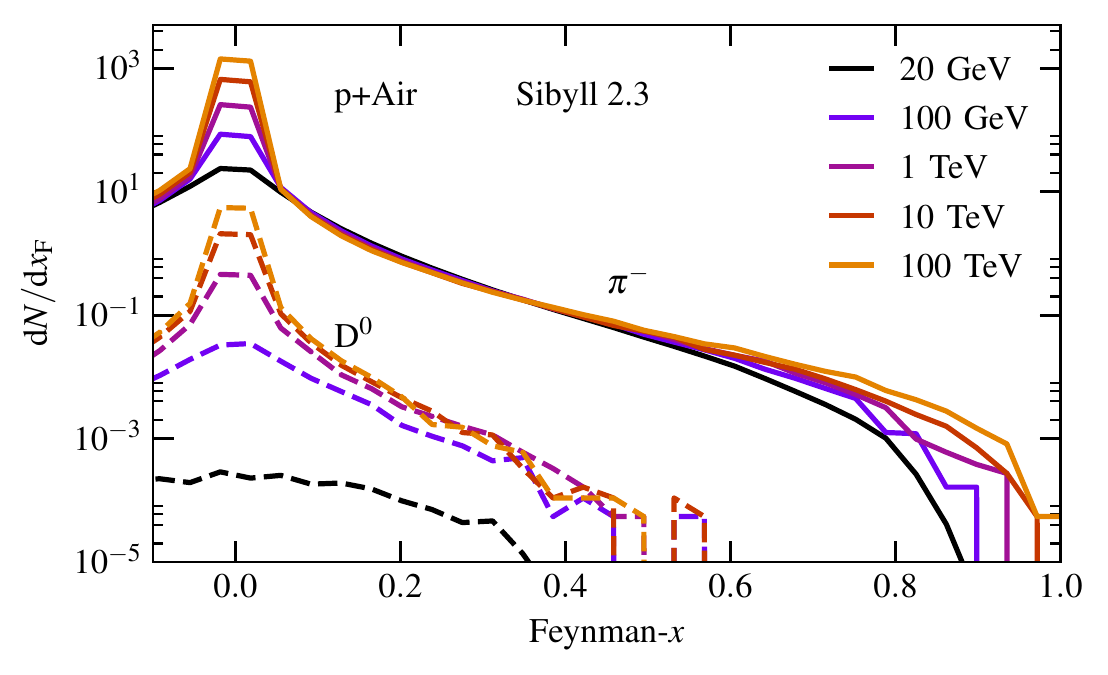}
  \caption{\label{fig:pion-spec-sib23} Spectrum of longitudinal
    momentum of $\pi^-$ in Sibyll~2.1 (left) and Sibyll~2.3
    (right). The hardening of the spectrum in Sibyll~2.3 is contrary
    to the expectation from Feynman scaling. For charmed hadrons
    (Sibyll~2.3 only) the scaling violations are not as prominent.}
\end{figure*}

\noindent
As an example for Feynman scaling, the production spectrum of $\pi^-$
for different energies in Sibyll~2.1 is shown in
Fig.~\ref{fig:pion-spec-sib23}~(left). As the cm.\ energy increases, more and
more particles are produced in the so-called central region ($x_{\rm
  F} < 0.1$, large-$p_{\rm T}$), while the rest of the spectrum
remains unchanged. As previously mentioned, Feynman scaling is not
explicitly implemented in the model but is an emergent phenomenon. It
is a direct, albeit not necessarily evident prediction of the naive
parton model and factorization~\cite{Feynman:1969ej}.
Adding conservation of energy, the
increase in the central region is expected to result in some softening
of the forward spectra.


In models that predict explicit violation of Feynman scaling in the forward region, 
parton distributions of soft and hard processes are connected (i.e.\ breaking factorization)
by, for example, deriving both from a common initial
state~\cite{Ostapchenko:2016ytp}. Due to the connection of the forward region to
the central region in such models, the extrapolation is still well
constrained for the entire phase space, even without Feynman scaling.
Different assumptions are, however, possible and lead to different extrapolations to high energy.
With the current understanding and existing accelerator data of leading particle production 
it is not possible to distinguish between different scenarios.

In Fig.~\ref{fig:pion-spec-sib23}~(right), the production
spectrum for Sibyll~2.3 is shown. Contrary to the general expectation, the spectrum
is shifted towards larger momentum fractions the higher the interaction energies are.
While the effect is not very pronounced, it shows that an undesired interplay between 
different processes responsible for leading particle production takes place. 

At parton level, Feynman scaling emerges from the factorization of the different
interaction scales. In addition the hadronization model used in
Sibyll, which is the string fragmentation
model~\cite{Andersson:1983jt,Sjostrand:1987xj}, has the property of
preserving the scaling (assuming a constant string
tension). One of the key differences
between Sibyll~2.1 (scaling) and Sibyll~2.3 (no scaling) is the new
treatment of excited beam remnants. While in Sibyll~2.1 the leading
particles produced by the beam remnants are effectively included by
adjusting the fragmentation process for partons related to the
incoming hadrons (valence quarks), Sibyll~2.3 dynamically produces
excited states whose decay products then emerge as the leading
particles. Although this remnant model seems a good
candidate to explain the unexpected scaling violations
in the forward region in Sibyll~2.3,
no direct connection was found.

\begin{figure*}[tb!]
  \includegraphics[width=0.5\columnwidth]{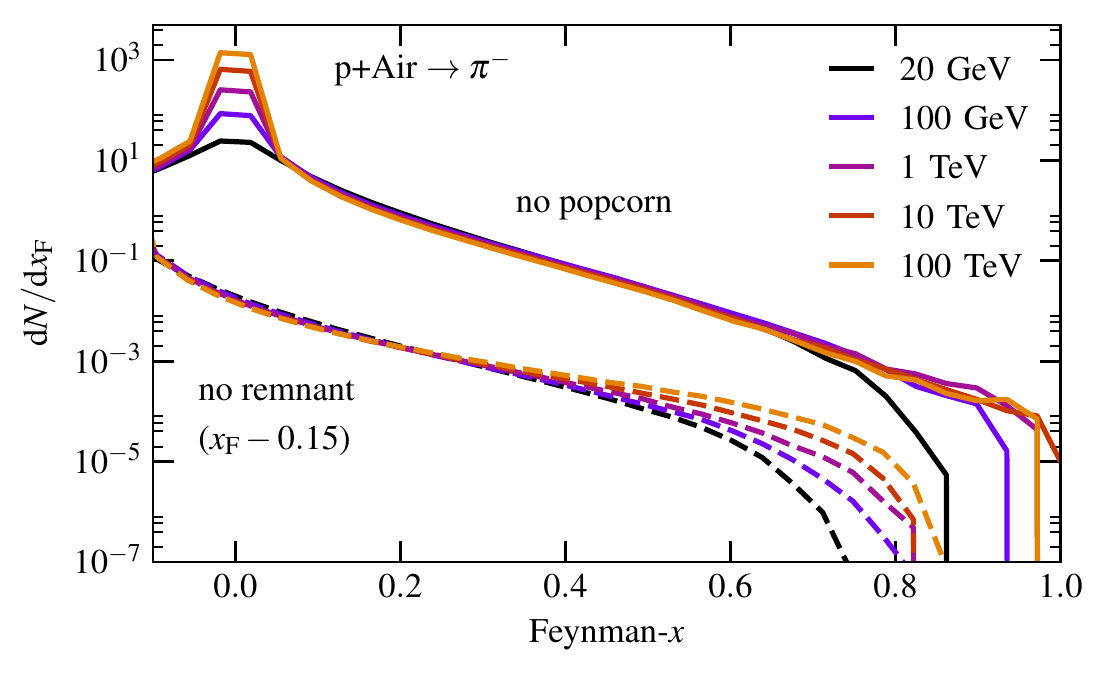}
  \hfill
  \includegraphics[width=0.5\columnwidth]{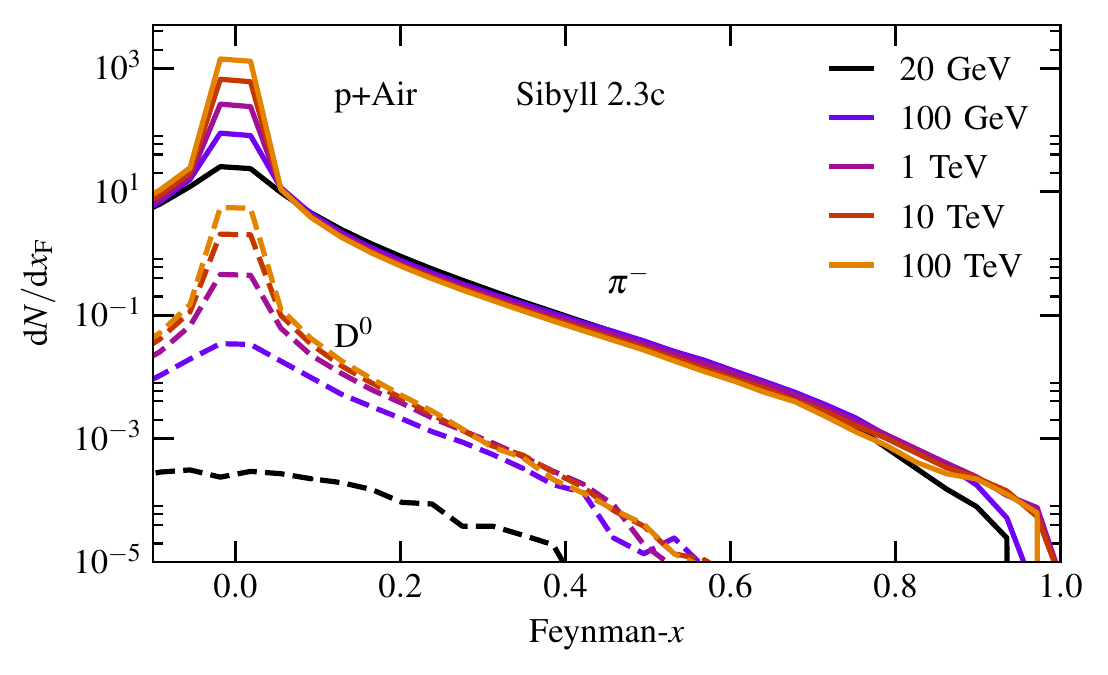}
  \caption{\label{fig:pion-spec-sib23c} Spectrum of longitudinal
    momentum of pions and charmed hadrons in Sibyll~2.3 without
    'popcorn' (left, solid lines), without the explicit remnant
    treatment (left, dashed lines) and in the retuned Sibyll~2.3c
    (right). The scaling violations in the fragmentation region in
    Sibyll~2.3 are induced by the interference of the popcorn
    mechanism with the leading quarks. The production of charmed
    hadrons in Sibyll~2.3c ($\mathrm{D}^0$, dashed lines on the right)
    is mostly unaffected because their production is dominated by
    central production.}
\end{figure*}

Instead, the violations were found to be induced by an extension to
string fragmentation that allows for the break-up of diquarks, the
so-called popcorn model~\cite{Andersson:1984af}. This model is used to
reduce the correlation between baryon-antibaryon pairs in the
fragmentation process, as observed in $e^+e^-$ annihilation. It also
leads to a broader rapidity distribution and
harder $x_{\rm F}$-spectrum. In combination with the fast diquark
from an initial proton it results in a significant hardening of the meson
spectra. In Fig.~\ref{fig:pion-spec-sib23c}~(left) the
restoration of the scaling behavior for Sibyll~2.3, if diquark
break-up disabled, is demonstrated. The effect of reverting the
explicit treatment of the remnant excitation to the effective
treatment in Sibyll~2.1 is also shown. The large deviation from
scaling at $20\,$GeV is a low-energy effect. It is related to 
the transition from string fragmentation to resonance production and decay.
The popcorn mechanism was
initially introduced to the model to improve this transition. In Sibyll~2.3c
this is compensated by allowing larger remnant excitation masses and by
retuning the fragmentation function and string tension. Furthermore, at low
energy, the so-called associated production $p^{\star}\to
\Lambda^0+\mathrm{K}^+$ was enhanced by adding it as an explicit final
state of the isotropic phase space decay (fireball).

The resulting spectrum is shown in
Fig.~\ref{fig:pion-spec-sib23c}~(right). Scaling behavior in the forward
region is successfully restored.

\subsection{Kaon spectra and atmospheric lepton fluxes}

\begin{figure}[tb!]
  \includegraphics[width=0.5\columnwidth]{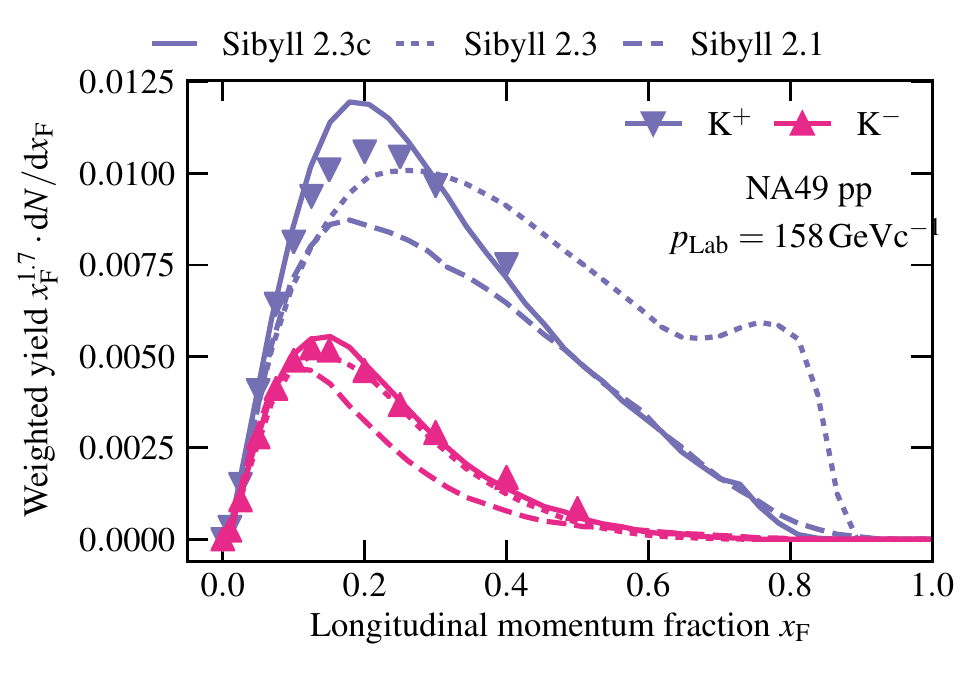}
  \hfill
  \includegraphics[width=0.5\columnwidth]{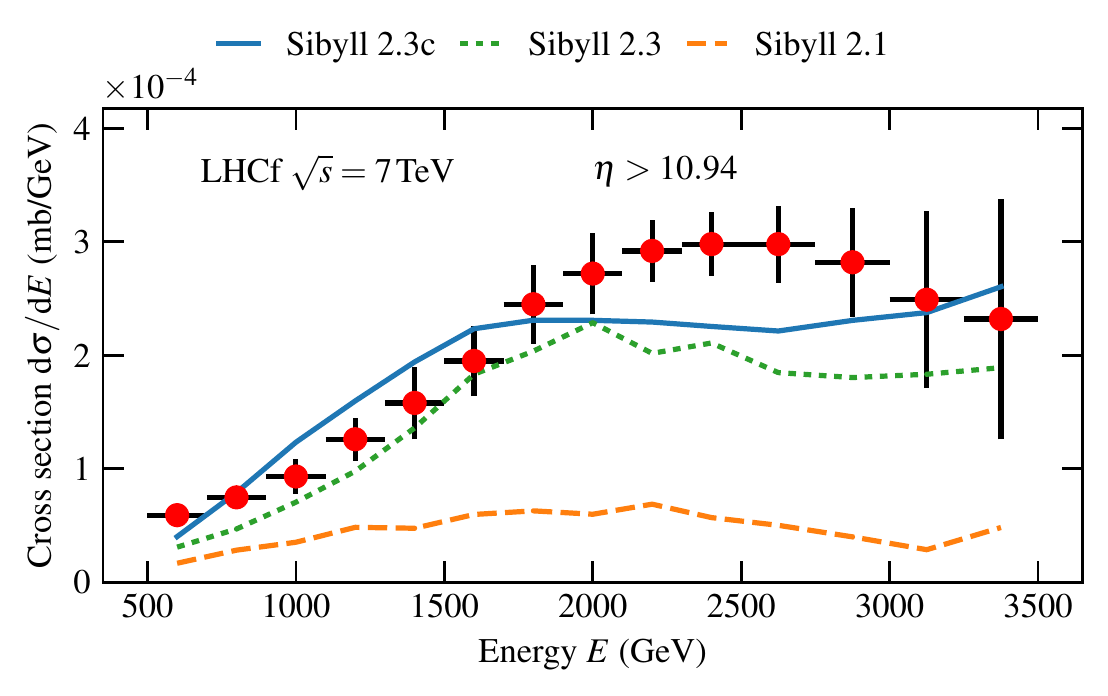}
  \caption{\label{fig:na49-kaons} Left: Weighted spectra of kaon
    production in pp interactions~\cite{Anticic:2010yg}. Right: Energy
    spectrum of neutrons at small angles~\cite{Adriani:2015nwa}. }
\end{figure}

\noindent
In addition to the interpretation of cosmic ray measurements, EAS simulations
are equally important in the context of high energy neutrinos, since
the bulk of cosmic rays produces a constant background flux of
atmospheric neutrinos. Integration over the whole
spectrum of cosmic rays shifts the weight of particle production into 
the forward fragmentation region~\cite{Gaisser:2002jj}. The hardening
of the longitudinal spectrum in Sibyll~2.3, therefore, has an even
larger impact on the predictions of atmospheric fluxes of muons and
neutrinos than its impact on EAS. In this context, kaons are of particular
importance.  Because of their shorter lifetime combined with the steep
cosmic-ray spectrum, their contribution to the
fluxes of atmospheric leptons increases above a $1\,$TeV.  The effect
is most important for $\nu_\mu$ where, because of the decay kinematics,
kaons become the dominant parent.  Figure~\ref{fig:na49-kaons}~(left)
illustrates the problem for Sibyll~2.3 by comparing its predictions separately
for production of K$^+$ and K$^-$ to the kaon spectra measured by
NA49~\cite{Anticic:2010yg}.  Production of K$^+$ is overestimated in Sibyll~2.3. 
(Note that a
weight factor of $x_{\rm F}^{1.7}$ is applied to reflect the
effect of the steep primary cosmic-ray spectrum.) The hard
spectrum of K$^+$ is due to the same effect as the
scaling violations discussed in the previous section, namely the
promotion of mesons to leading particles by diquark break-up. This
interpretation is confirmed by noting that the effect is not present
for negatively charged kaons. This difference between the charge
states is expected given the absence of strange valence quarks 
in the proton, so the negative charge state (K$^-(\bar{u}s)$) cannot
be associated with leading quarks. With the adjustments\footnote{
The hadronization parameters for the updated Sibyll~2.3c were found by
fitting a parameterization of the event generator response to the pp
measurements by NA49 using the Professor~\cite{Buckley:2009bj} and
Rivet~\cite{Buckley:2010ar} tools.}
in Sibyll~2.3c,
the shape of the kaon spectrum is reproduced more accurately. At the same time the
advantages of the explicit treatment of the remnant excitations are
preserved, as is shown in the comparison with the energy spectrum of
neutrons measured by LHCf~\cite{Adriani:2015nwa} in
Fig.~\ref{fig:na49-kaons}~(right).

The predictions of Sibyll~2.3c for the atmospheric lepton
fluxes are presented in a separate contribution at this conference~\cite{Fedynitch:2017xx1}.

\section{Predictions for extensive air showers}

\noindent
The update of the hadronic interaction model
Sibyll presented here mainly concerns the evolution of the shape of the
production spectra of mesons at large values of Feynman-$x$. Due to
the weighting with the primary spectrum in the calculation of the
atmospheric fluxes of leptons these changes at large
$x_{\rm F}$ will have a larger effect on the atmospheric fluxes than
on air shower predictions, where the production spectra enter via $x
\, \mathrm{d}\sigma /\mathrm{d}x$. In addition, kaons play a smaller
role in air showers. 

In the following we show predictions calculated with
CONEX~\cite{Bergmann:2006yz}. The energy threshold for
the transition between the Monte Carlo and numerical cascade equation
was set to 5\% of the primary energy.

\begin{figure*}[tb!]
  \includegraphics[width=0.5\columnwidth]{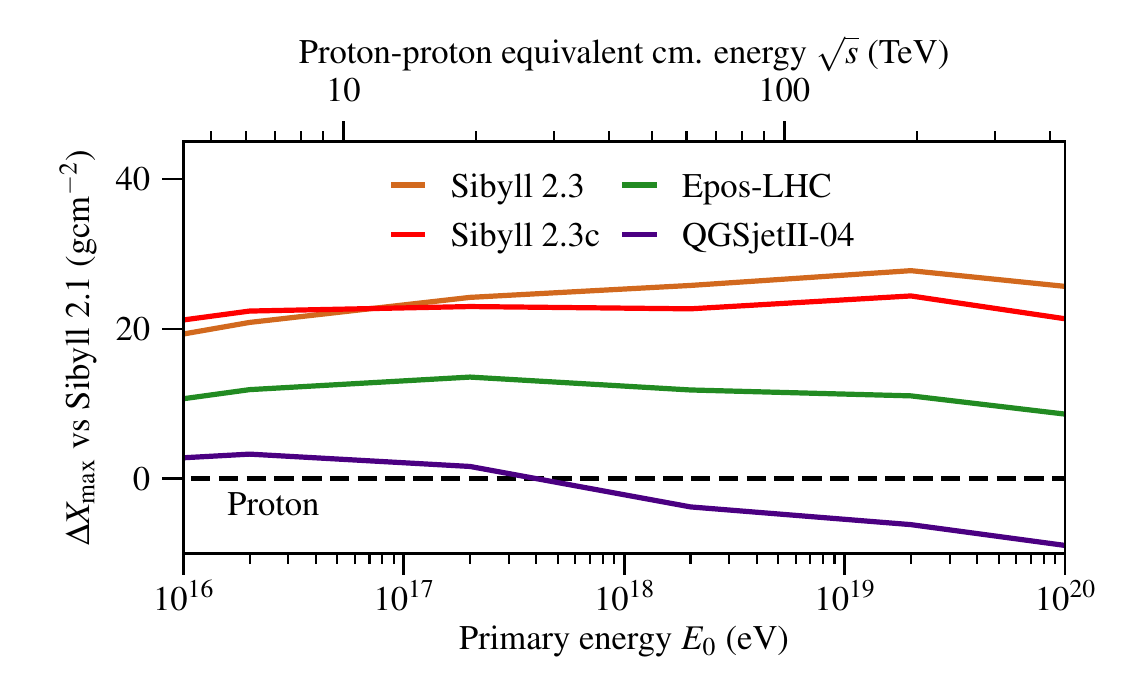}
  \hfill
  \includegraphics[width=0.5\columnwidth]{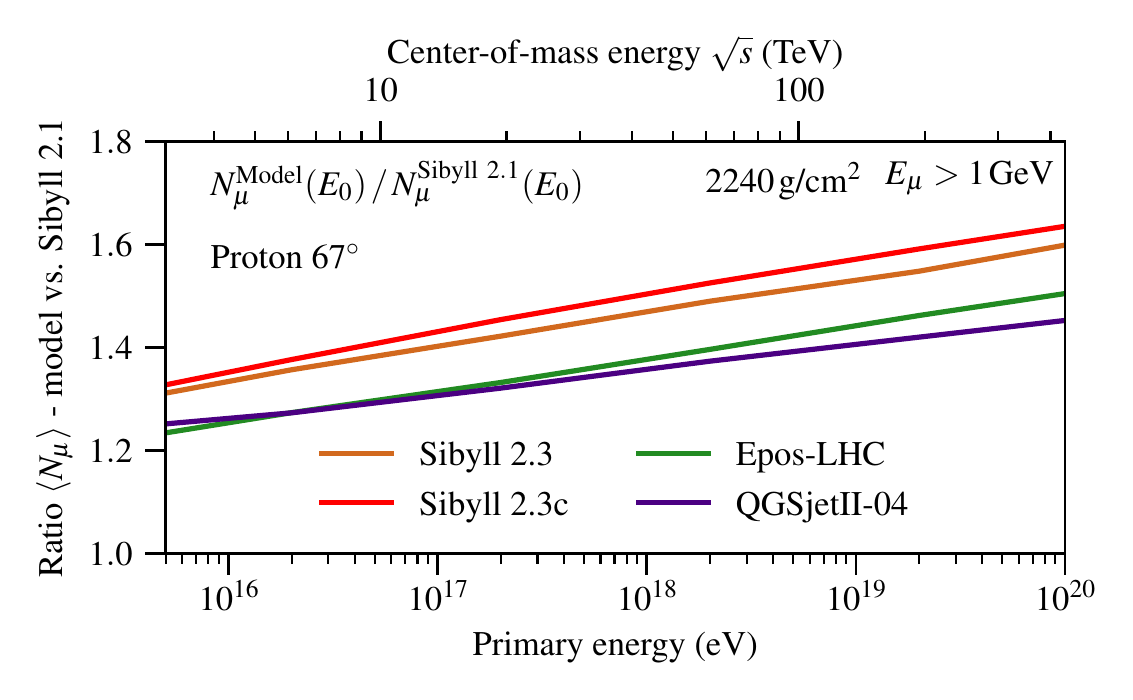}
  \caption{\label{fig:eas-predict} Difference between post-LHC models
    and Sibyll~2.1 for \xmax (left) and \nmu (right).}
\end{figure*}

For air showers, the updated model Sibyll~2.3c predicts a slightly
shallower depth of the average position of shower maximum (\xmax) than
Sibyll~2.3. Compared to Sibyll~2.1, \xmax is deeper by
$20\,$g$/$cm$^2$ across the energy range of ultra-high-energy cosmic rays, see
Fig.~\ref{fig:eas-predict}~(left). The number of muons (\nmu) is
larger than in Sibyll~2.3, albeit only by a very small amount
(Fig.~\ref{fig:eas-predict}~(right)). Relative to Sibyll~2.1, \nmu is
larger by a factor $\sim 1.35$ at $10^{16}\,$eV and by a factor $\sim
1.6$ at $10^{20}\,$eV. In the current version, Sibyll predicts the
largest number of muons of all post-LHC interaction
models~\cite{Pierog:2013ria,Ostapchenko:2010vb}. There are no new
processes in Sibyll~2.3c wrt.\ Sibyll 2.3 that could cause the slight 
increase in muon number,
the only change is the shape of the production spectra.

\begin{figure*}[tb!]
  \includegraphics[width=0.5\columnwidth]{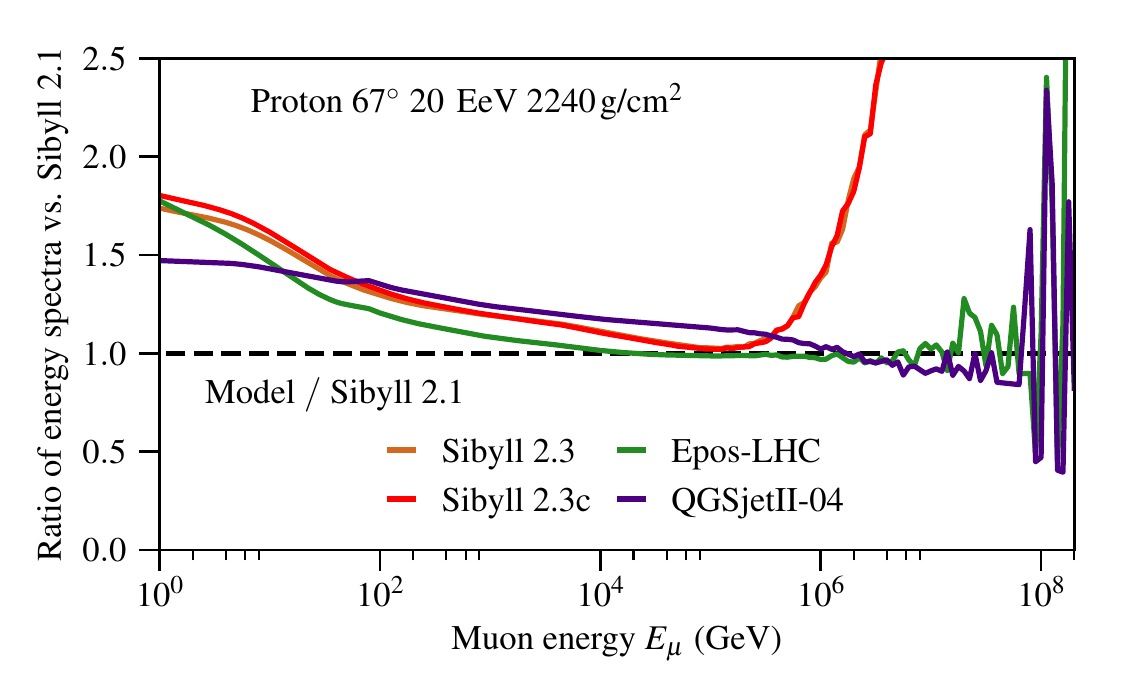}
  \hfill
  \includegraphics[width=0.5\columnwidth]{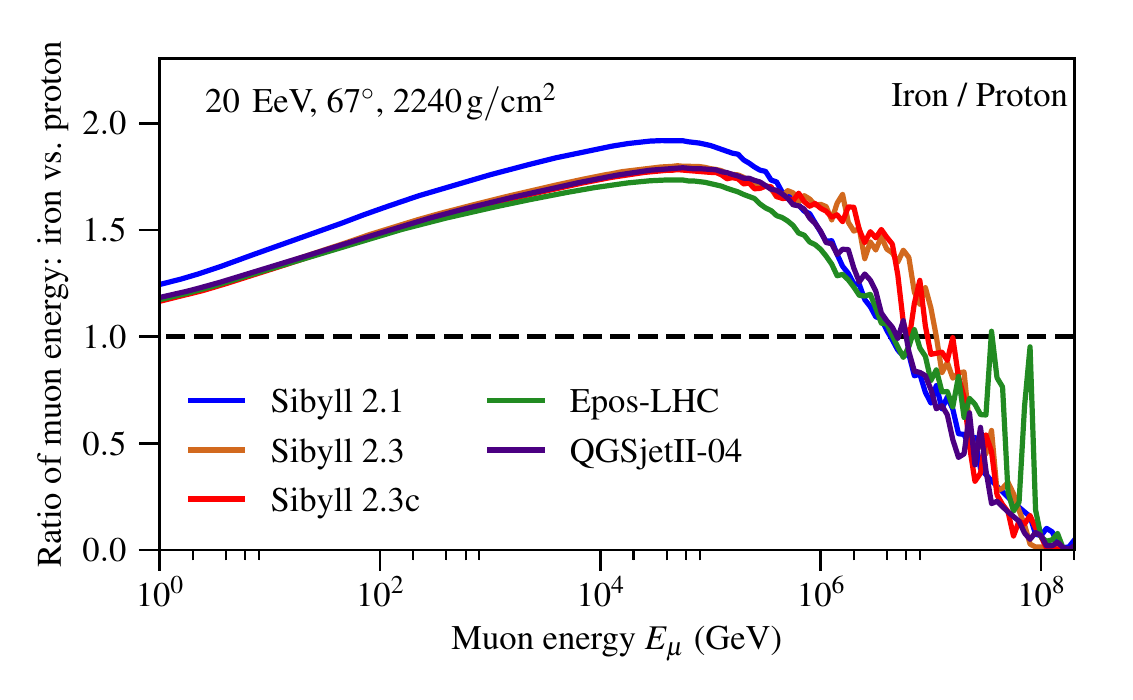}
  \caption{\label{fig:eas-mu-spec} Ratio between the energy spectrum
    of muons for the post-LHC models and Sibyll~2.1 (left) and between
    different primaries (right). The primary energy is $10^{19}\,$eV
    and the spectra are evaluated at a depth of $2240\,$g$/$cm$^2$.}
\end{figure*}

The effect on muon production is summarized by the change in the
energy spectrum of muons (Fig.~\ref{fig:eas-mu-spec}). The figure
shows that the softening of the spectra and the restoration of Feynman
scaling leads to the increase of muons with low-energy. Below
$100\,$GeV Sibyll~2.3c produces most muons of all models. Beyond
$1\,$PeV the contribution from the prompt decays of charmed hadrons
sets in. The strong increase of the number of muons at low energies of
the post-LHC models relative to Sibyll~2.1 is due to: (i) increased
production of baryon-antibaryon pairs, (ii) the increased ratio
$\pi^{\pm}:\pi^0$ in pion interactions in forward direction, mainly stemming from the
formation of leading $\rho^0$ resonances instead of leading $\pi^0$ mesons.

In Fig.~\ref{fig:eas-mu-spec}~(right) the ratio of the energy
spectrum of muons of iron and proton showers is shown. With
the post-LHC models, the overall difference between proton and iron
decreases. In contrast to the energy spectrum of muons for the
individual nuclei, the models agree very well on the difference
between nuclei, in particular for muons of low energy.

\section{Conclusion \& Outlook}

\noindent
We have developed a new version of Sibyll, called Sibyll~2.3c, by re-tuning the 
model version 2.3 to obtain a better description of NA49 data.
It was found that Sibyll~2.3c approximately obeys Feynman scaling in the
fragmentation region up to the highest energies and gives a
better description of the measured kaon production spectra than
previous versions. The predictions for extensive air showers
are very similar to Sibyll~2.3, as are the predictions for
production of charmed hadrons. Larger changes are found for inclusive
fluxes of atmospheric leptons, which are discussed in~\cite{Fedynitch:2017xx1}.


\paragraph{Acknowledgements}
We thank our colleagues T.~Pierog and S.~Ostapchenko for many fruitful
discussions and constructive criticism. This work is supported in part
by the U.S.\ National Science Foundation (PHY-1505990) and in part by
the KIT graduate school KSETA.

\bibliographystyle{JHEParxivfix}
{\sloppy
  \bibliography{references,local}
}


\end{document}